\documentclass[twocolumn,pra]{revtex4}
\usepackage[latin9]{inputenc}
\usepackage{color}
\usepackage{amsmath}
\usepackage{amssymb}
\usepackage{graphicx}

\makeatletter
\@ifundefined{textcolor}{}
{%
 \definecolor{BLACK}{gray}{0}
 \definecolor{WHITE}{gray}{1}
 \definecolor{RED}{rgb}{1,0,0}
 \definecolor{GREEN}{rgb}{0,1,0}
 \definecolor{BLUE}{rgb}{0,0,1}
 \definecolor{CYAN}{cmyk}{1,0,0,0}
 \definecolor{MAGENTA}{cmyk}{0,1,0,0}
 \definecolor{YELLOW}{cmyk}{0,0,1,0}
}

\usepackage{amsfonts}
\newtheorem{theorem}{Theorem}
\newtheorem{acknowledgement}[theorem]{Acknowledgement}

\makeatother

\begin{document}

\title{{\Large{}Frequency stabilization and tuning of breathing soliton
in SiN microresonators}}

\author{Shuai Wan$^{1,3}$}
\author{Rui Niu$^{1,3}$}
\author{Zheng-Yu Wang$^{1,3}$}
\author{Jin-Lan Peng$^{2}$}
\author{Ming Li$^{1,3}$}
\author{Jin Li$^{1,3}$}
\author{Guang-Can Guo$^{1,3}$}
\author{Chang-Ling Zou$^{1,3}$}
\email{clzou321@ustc.edu.cn}
\author{Chun-Hua Dong$^{1,3}$}
\email{chunhua@ustc.edu.cn}

\affiliation{$^{1}$CAS Key Laboratory of Quantum Information, University of Science
and Technology of China, Hefei, Anhui 230026, P. R. China.}

\affiliation{$^{2}$Center for Micro and Nanoscale Research and Fabrication, University
of Science and Technology of China, Chinese Academy of Sciences, Hefei
230026, P. R. China.}

\affiliation{$^{3}$CAS Center For Excellence in Quantum Information and Quantum
Physics, University of Science and Technology of China, Hefei, Anhui
230026, P. R. China.}

\begin{abstract}
Dissipative Kerr soliton offers broadband coherent and low-noise frequency
comb and stable temporal pulse train, having shown great potential
applications in spectroscopy, communications, and metrology. Breathing
soliton is a particular dissipative Kerr soliton that the pulse duration
and peak intensity show periodic oscillation. However, the noise and
stability of the breathing soliton is still remaining unexplored,
which would be the main obstacle for future applications of breathing
solitons. Here, we have investigated the breathing dissipative Kerr
solitons in the silicon nitride (SiN)microrings, while the breather
period shows uncertainties around MHz in both simulation and experiments.
By applying a modulated pump, the breathing frequency can be injectively
locked to the modulation and tuned over tens of MHz with frequency
noise significantly suppressed. Our demonstration offers an alternative
knob for the controlling of soliton dynamics in microresonator and
paves a new avenue towards practical applications of breathing soliton.

\end{abstract}
\maketitle

\section{INTRODUCTION}

Optical solitons which maintain their localized structures during
propagation can be generated by balancing dispersion and nonlinearity
in the propagation media \cite{K-review2018,soliton-2}. This phenomenon
was firstly demonstrated in optical fiber, aiming to transmit information
with increased bandwidth \cite{soliton-3}. Since then, this kind
of localized pulses has been explored and realized in various nonlinear
systems and enables a variety of scientific and technological applications
\cite{soliton-4,soliton-6,soliton-5,xiaoxiao,Zengheping}.  Whispering
gallery mode (WGM) microresonators, hold the advantages of high optical
quality factor (Q-factor) and the small mode volume \cite{V2003},
could greatly enhance light-matter interaction and thus provide an
alternative platform for studying these fascinating nonlinear physics~\cite{soliton-6,soliton-5,K-soliton2014,wong}.
In addition, the microresonator in the integrated photonic platform
is excellent for future applications with high stability, reduced
size and costs.

\begin{figure}
\centerline{\includegraphics[clip,width=0.9\columnwidth]{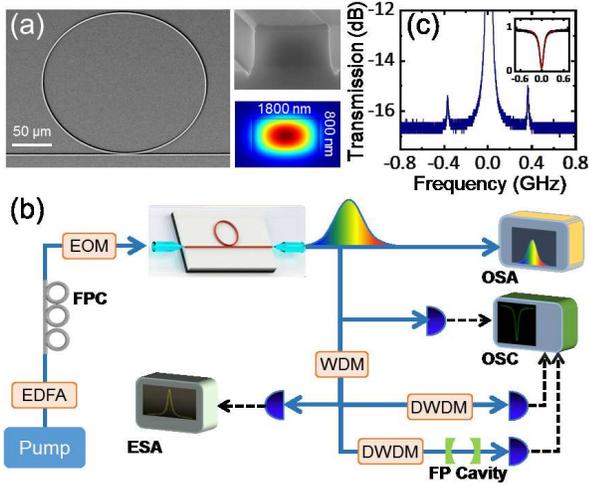} }\caption{\label{fig:Fig1} (a) Scanning electron micrographs of $\mathrm{Si_{3}N_{4}}$
microring with diameter of $200\,\mathrm{\mu m}$. Insets show the
microring cross-section of $1.8\,\mathrm{\mu m}\times0.8\,\mathrm{\mu m}$
and the corresponding fundamental transverse-magnetic mode profile. (b) The experimental setup for Kerr frequency comb
generation. EDFA, FPC, EOM, WDM, DWDM, FP, OSC, and ESA are erbium-doped
fiber amplifier, fiber polarization controller, electro-optical modulator,
wavelength division multiplexing, dense wavelength division multiplexing,
Fabry-Perot cavity, oscilloscope and electronic spectrum analyzer,
respectively. (c) Detailed comb line spectrum of a breathing microsoliton
measured by the FP spectrum analyzer, with two sidebands indicating
the breathing frequency around $0.4\,\mathrm{GHz}$. The inset shows
a typical resonance of the microring, with a loaded Q of $1.5\times10^{6}$
according to the Lorentzian fitting (red line).}
\vspace{-6pt}
\end{figure}

The first observation of optical microcomb was realized in 2007~\cite{K-firstcomb}
and after several years of deep studies in overcoming challenges of
low coherence, optical soliton was successfully generated in a crystalline
microresonator~\cite{K-soliton2014}. Compared to the aforementioned
soliton generation schemes, in addition to the balance of dispersion
and nonlinearity, soliton generation in WGM microresonators also requires
the balance of loss and gain. Such optical soliton is called dissipative
Kerr soliton (DKS) and firstly observed in the mode-locked fiber laser~\cite{DKS-fiber}.
Different from the mode-locked fiber laser, where the gain is from
an active media, the gain in the passive WGM microresonator is the parametric gain of four-wave mixing (FWM) stimulated by an external
continuous-wave (CW) laser. The rapid development of DKS in microresonators
makes it quite a good candidate for applications in ultrahigh data
rate communication~\cite{K-comm2017}, quantum key distribution~\cite{QKD},
high precision optical ranging~\cite{V-rangemea2018}, dual-comb
spectroscopy \cite{V-dualcomb2016,L-dualcomb2018}, low-noise microwave
source~\cite{K-microwave2019}, optical clock~\cite{NIST-clock2019}
and astronomical spectrometer calibration~\cite{V-astrocomb2019,K-astrocomb2019}.

Associate with DKS, a series of novel nonlinear phenomena have been
observed, such as dark pulses, Cherenkov radiation, Raman self-frequency
shift, and breathing soliton. In particular, breathing solitons exhibit
periodic oscillation behavior in both amplitude and pulse duration,
which is related to the Fermi-Pasta-Ulam recurrence~\cite{Weiner-breather2016}.
It has drawn a lot of attentions for fundamental studies of nonlinear
physics and was demonstrated in experiments recently \cite{Weiner-breather2016,L-breather2017,K-breather2017,K-intermode2017,Weiner-darkbreather2018}.
It is revealed that breather soliton state could exist in the detuning
region between modulation instability state and stable DKS state because
of intrinsic dynamical instability \cite{Weiner-breather2016,L-breather2017,K-breather2017}
or in the conventionally stationary DKS region because of the intermode
interaction like avoided mode crossing \cite{K-intermode2017}. Furthermore,
in addition to bright breather solitons, dark breather solitons have
also been observed in normal dispersion microresonators \cite{Weiner-darkbreather2018}.
However, in the aforementioned studies, only the basic features of
the breathing behavior are reported, and the breathing soliton is
treated as unwanted because of the degradation of soliton stability.
Therefore, the stability or noise properties of the breather soliton
are not carefully studied, and the potential applications of the breath
soliton are still remaining unexplored.

In this work, we demonstrate the stabilization and controlling of
the breathing frequency of microsoliton, for the first time among
all platforms, via injection locking. Through both numerical and experimental
studies, we demonstrate that the breathing frequency can be injection-locked
by applying an appropriate modulation signal to pump laser, and then
the phase noise of the system is remarkably suppressed. Furthermore,
the stabilized breathing frequency is tuned over ${\color{red}50}\,$MHz.
Our approach to stabilize and control the breathing frequency paves
the way towards the applications of the breathing soliton. For example,
the breathing soliton provides a robust way to transfer RF frequencies
between very different optical frequency lines through broadband microsoliton.
Additionally, our demonstration also provides a convenient approach
to study the differential absorption spectra, by comparing the sidebands
of each comb line due to the locked breathing frequency modulation.

\begin{figure}
\centerline{\includegraphics[clip,width=0.9\columnwidth]{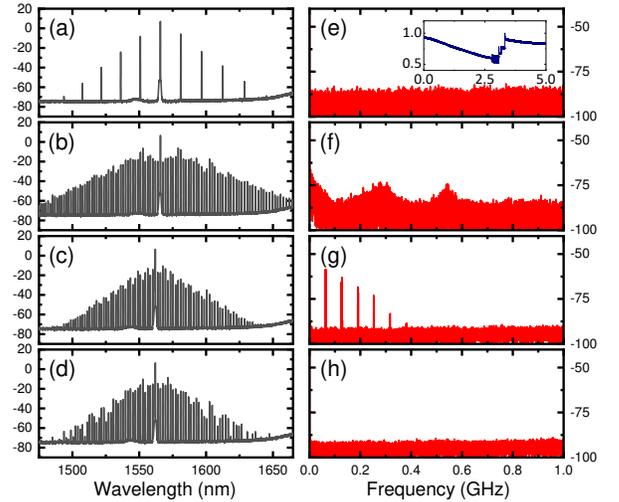}}\caption{The evolution of the microsoliton generation processes during the
scanning of the pump laser detuning. (a-d) Typical optical spectra.
Four evolution stages are primary comb (a), modulation instability
comb (b), breather soliton (c) and stable soliton (d), respectively.
(e-h) The corresponding evolution of RF spectra. \textcolor{black}{Inset:
The transmission spectrum of the microring when the laser frequency
is scanning across the resonance mode.}}
\label{fig:Fig2}\vspace{-6pt}
\end{figure}

\section{Device fabrication and setup of soliton generation}

In our experiment, $\mathrm{Si_{3}N_{4}}$ integrated microring resonator
is fabricated with a diameter of $200\,\mu m$ and a cross-section
of $1.8\,\mathrm{\mu m}\times0.8\,\mathrm{\mu m}$, which is designed
for the anomalous group velocity dispersion that required for frequency
comb generation. Figure~\ref{fig:Fig1}\textcolor{black}{(a) }shows
the scanning electron microscopy (SEM) photon of a typical device,
which consists of a microring and a straight bus waveguide. The SEM
of the microring cross-section before removing the residual photoresist
is shown in the upper right inset, with the corresponding fundamental
optical mode profile shown in the bottom inset. The devices are fabricated
from a silicon substrate wafer with $500\,\mathrm{\mu m}$ silicon,
$3\,\mathrm{\mu m}$ wet oxidation silicon dioxide ($\mathrm{SiO_{2}}$)
and 800~nm deposited stoichiometric $\mathrm{Si_{3}N_{4}}$. To prevent
cracks by high tensile stress of the $\mathrm{Si_{3}N_{4}}$ film,
some trenches are created around the wafer by a diamond scribe before
the deposition \cite{L-fab2013}. A series of optimization techniques
are also performed during the fabrication processing and finally,
an upper cladding of $3\,\mathrm{\mu m}$ $\mathrm{SiO_{2}}$ is deposited
by using plasma-enhanced chemical vapor deposition (PECVD) to cover
the sample. More details about the fabrication process are provided
in the Supporting Information. Then, the devices are tested by the
setup shown in Fig.~\ref{fig:Fig1}(b). A tunable CW laser (Toptica
CTL 1550) is coupled into and out of the chip through lensed fibers.
The coupling loss of lensed fibers is measured to be $\sim3.5\,\mathrm{dB}$
at each facet.  In the inset of Fig.~\ref{fig:Fig1}(c), a typical
transmission spectrum of a fundamental mode is shown and fitted by
a loaded $Q$-factor of $\sim1.5\times10^{6}$. The high extinction
ratio of the resonance indicates a high intrinsic $Q$-factor of $3\times10^{6}$
and high efficiency energy delivering from the bus waveguide to the
modes, both are beneficial for nonlinear optics effects.

Soliton generation is achieved by sweeping the pump laser frequency
over a resonance mode with higher power of $100$ mW, from the blue-detuned
regime to the red-detuned regime~\cite{K-review2018,wong,K-comm2017,L-dualcomb2018}.
During the laser frequency scanning, the cavity fields build up, and
strong nonlinear four-wave mixing effect induces the comb generation
and essentially realizes the soliton. A complete spectral evolution
can be divided into four stages, including primary comb, modulation
instability (MI) comb, breather soliton, and stable soliton, as shown
by the optical spectra and radio frequency (RF) spectra in Fig.~\ref{fig:Fig2}.
The inset is the corresponding transmission spectrum when scanning
the pump. By carefully monitoring the transmitted power and stopping
the pump frequency scanning at a certain detuning point, the four
stages could be repetitively accessed in experiment.

\begin{figure}
\centerline{\includegraphics[clip,width=0.9\columnwidth]{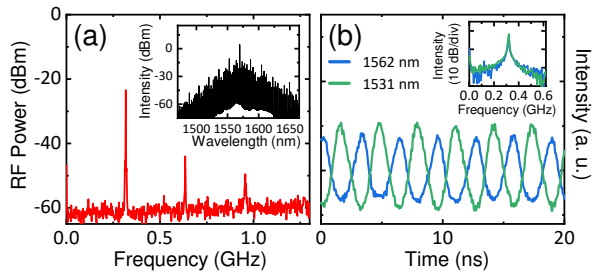}}\caption{Features of breathing soliton. (a) The detailed RF spectrum of a breather
soliton state. Inset: the corresponding optical spectrum. (b) The
recorded fast power evolution of a single comb line around the center
($1562\:nm$, blue curve) and in the wings ($1531\:nm$, green curve)
of breather optical spectrum, respectively. Inset: the corresponding
Fourier transform spectrum.}

\label{fig:Fig3}\vspace{-6pt}
\end{figure}

\section{Results and discussions}

\subsection{Breathing soliton}

Among the four stages of the comb evolution, we are particularly interested
in the breathing soliton. We verified the generation of breathing
solitons in different devices with varying pump conditions. Figure
~\ref{fig:Fig1}(c) shows the optical spectrum measured by a Fabry-Perot
cavity spectrometer. Two sidebands are obvious around
the comb line, showing a breathing frequency $f_{\mathrm{br}}$ of
about $380\:\textrm{MHz}$. In another device, we studied the RF spectrum
and temporal oscillation of individual comb lines, as shown in Fig.~\ref{fig:Fig3}.
From the RF spectrum, we can find that the breathing frequency around
$320\:\textrm{MHz}$. Then, the dense wavelength division multiplexing
(DWDM) is used to filter out the single comb line around the center
(\textasciitilde$1562\:\mathrm{nm}$ ) and the wings (\textasciitilde$1531\:\mathrm{nm}$ ),
which is detected by a fast photodetector (PD) and recorded by an
oscilloscope, as shown in Fig.~\ref{fig:Fig3}(b). Here, we record the traces over $200\:\textrm{ns}$ and compute the RF spectrum based on the Fourier tranform. As shown in the inset of of Fig.~\ref{fig:Fig3}(b), the oscillation
frequency is same as the frequency shown in Fig.~\ref{fig:Fig3}(a).
And the traces of the two comb lines are nearly out-of-phase when
compared with each other, revealing that there exists a periodic energy
exchange between comb lines around the center and the wings, which
is a characteristic of Fermi-Pasta-Ulam recurrence and a typical signature
of breather solitons.

Comparing the breathing solitons in different experiments, we found
that the breathing behavior could be robustly obtained in our devices,
while its frequency varies from device to device. Even for a single
device, the breathing frequency is not fixed. Here,
we observe that the linewidth of the RF beatnote is on the order of
megahertz, indicating a fast drifting of the breathing frequency.
In the time traces, the fluctuations are obvious even in $10\,\mathrm{ns}$
time-scale. Therefore, the problems of the inhomogeneous breathing
frequencies and the frequency instability hinder the potential applications
of the breathing soliton.

\subsection{Scheme of injection-locking}

\begin{figure}[b]
\centerline{\includegraphics[clip,width=1\columnwidth]{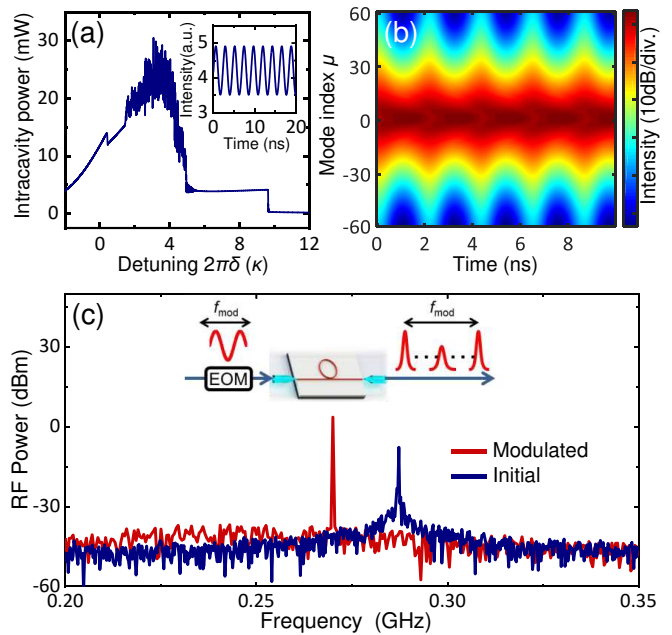}}\caption{(a) The simulated evolution of the intracavity power when the laser
frequency is scanning across the resonance mode. The inset shows the
oscillations of the power for a fixed laser frequency in the breather
soliton state. (b) Periodic spectrum evolution of a breather soliton
state. (c) RF spectra of initial breather soliton state (blue line)
and modulated breather soliton state (red line). The initial breathing
frequency $f_{br}$ is $287\:\mathrm{MHz}$ and the modulated frequency
$f_{mod}$ is $270\:\mathrm{MHz}$. Inset shows the concept of injetcion
locking of breather soliton. A modulation signal with $f_{mod}$ is
applied to the pump laser after the appearance of breather soliton
and $f_{br}$ is injection locked if $f_{mod}$ is within the locking
range.}
\label{fig:Fig4}\vspace{-6pt}
\end{figure}

The stabilization and controlling of breathing frequency is of great
significance, because it is not only helpful for a better understanding
of nonlinear dynamics in microresonators, but also critical for many
potential applications. Here, we envision two intriguing application
scenarios enabled by the breathing soliton. First, we could use a
single breather soliton, instead of two microresonators in the dual-comb
scheme, for applications in spectroscopy. As can be noted from Fig.~\ref{fig:Fig1}c,
with the occurrence of the breather soliton, a series of sidebands
can be observed around each comb line. Compared with the mode spacing
of each comb line, these sidebands have much smaller frequency separation,
which is equal to the breathing frequency. Therefore, the generated
sidebands can provide additional differential absorption spectra around
each comb line and an enhanced local resolution is expected. Second,
the breathing could transfer a sub-GHz signal among all comb lines
that would span an octave of wavelength. Thus, the breathing soliton
could be used for distributing RF frequency references.

However, from our experiments, the breathing frequency depends on
many practical parameters, such as the pump power, pump detuning,
temperature of the chip, the polarization in the fiber and the coupling between the fiber lens and the
chip. It is almost impossible to
suppress all these imperfections in the experiments. Therefore,
we propose to lock the breathing frequency by an external injection
of RF signal, as illustrated in the inset of Fig.~\ref{fig:Fig4}(c).
An intensity electro-optic modulator (EOM) was introduced to modulate
the pump power with an appropriate modulation signal. The frequency
sidebands generated by the modulator could stimulate the four-wave
mixing in a single mode and expand mode to mode during soliton generation.
This stimulated process competes with the natural oscillation induced
by spontaneous four-wave mixing in breather soliton. We expect it
dominates the sidebands generation with strong enough modulation depth, the wide oscillation spectrum caused by spontaneous process and
environment fluctuation be suppressed.

In order to investigate the feasibility of this approach, the numerical
simulation is performed based on the coupled-mode equations (CME)
\cite{CME-Chembo-2010,CME-2014,CME-Tang-2018}. First of all, we numerically
test the stages of frequency comb generation with fixed pump power
of $100\:mW$. The simulation parameters of the microring are the
same as the device used in our experiments. Figure~\ref{fig:Fig4}a
shows the simulated evolution of the intracavity power when scanning
the pump frequency. For more details about the numerical model, see
Supplementary materials. In good agreement with our experiments {[}Fig.~\ref{fig:Fig2}{]},
there are clearly four states as the laser is tuning across the resonance
mode~\cite{soliton-5}. By stopping the laser frequency on the breather
soliton stage, it can be observed that the intracavity power experiences
the periodic oscillation, as shown in the inset of Fig.~\ref{fig:Fig4}(a).
A spectral envelope evolution of the breather soliton is also depicted
in Fig.~\ref{fig:Fig4}(b), which shows the power of the breather
soliton flows to the wings of the spectrum and returns to the center
periodically, agrees with Fig.~\ref{fig:Fig3}(b). Then, the injection
locking scheme is numerically investigated by applying a modulation
signal on the pump power with a modulation frequency $f_{\mathrm{mod}}$,
which is close to $f_{\mathrm{br}}$. Figure~\ref{fig:Fig4}(c) shows
the RF spectra for the breathing soliton with and without injection-locking.
It is found that the breathing frequency is locked to the $f_{\mathrm{mod}}=270\:\mathrm{MHz}$
(red line) even it is about $17\,\mathrm{MHz}$ away from its natural
breathing frequency (\textasciitilde$287\,\mathrm{MHz}$ , blue line).
In addition, the RF noise background is suppressed by the injection
locking, and the linewidth of the beatnote is much narrower than without
injection locking.

\subsection{Stabilization of breathing frequency}

\begin{figure}
\centerline{\includegraphics[clip,width=1\columnwidth]{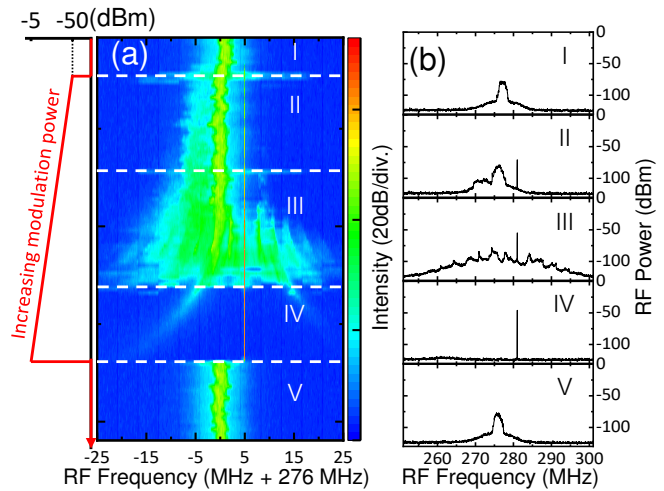}}\caption{\label{fig:Fig5}(a) Evolution of the RF spectrum when gradually increasing
the modulation power from -50dBm to -5dBm. The initial breathing frequency
$f_{br}$ is $276\:\mathrm{MHz}$ ($\text{I}$) and the modulation
frequency $f_{mod}$ is $281\:\mathrm{MHz}$. With the increase of
the modulation power, there is a competition between $f_{br}$ and
$f_{mod}$ and other harmonics components appears ($\text{II}$ and
$\text{III}$). Eventually, $f_{br}$ is synchronized to $f_{mod}$
as the modulation power is strong enough ($\text{IV}$). $f_{br}$
returns back to the initial frequency after turning off the modulation
signal (V). (b) Snapshots with different evolution stages in (a).}
\vspace{-6pt}
\end{figure}

With the scheme validated numerically, we performed the injection
locking for the breathing frequency stabilization. In our experiment,
we use a sinusoidal wave signal to drive the EOM and monitor the evolution
of the RF spectrum around the initial $f_{br}$ when the modulation
power varies from $-50\:\mathrm{dBm}$ to $-5\:\mathrm{dBm}$ with
frequency of $281\:\mathrm{MHz}$. The experimental results are summarized
in Fig.~\ref{fig:Fig5}. Before we turn on the modulator, a breather
soliton state was generated at a $f_{br}$ of $276\:\mathrm{MHz}$
and the linewidth of the beatnote on the RF spectrum is relatively
wide (stage $\text{I}$). When we input the modulation signal and
gradually increase the modulation power, as shown in Fig. 5a (stage
$\text{II}$), a sharp peak at $f_{\mathrm{mod}}$ appears in the
RF spectrum and the intensity of the $f_{br}$ beatnote becomes weak,
indicating that there exists a competition between this two frequencies
and they can coexist when the modulation power is not strong enough.
With further increasing the modulation power, $f_{mod}$ gradually
gains the upper hand in the competition and multiple harmonics components
arising from $f_{mod}$ and $f_{br}$ can be observed in the RF spectrum,
which raises the noise of the background (stage $\text{III}$) along
with the weak $f_{br}$ beatnote. Eventually, as the modulation power
is sufficiently strong, both the beatnote of $f_{br}$ and other harmonics
components vanish in the RF spectrum and only a beatnote of $f_{mod}$
with narrow linewidth exists (stage $\text{IV}$). According to the
recorded out-of-phase oscillatory power traces of comb lines around
the center and in the wings at this stage, we can confirm that the
breather soliton state still remains and $f_{br}$ is synchronized
to $f_{mod}$. Finally, we turn off the modulation signal and $f_{br}$
immediately returns to the initial $f_{br}$ and the linewidth became
as wide as before (stage $\text{V}$). Since our experiment performs
with a free-running pump laser without any stabilization scheme and
$f_{br}$ is sensitive to the pump power and effective detuning of
the pump laser, as shown in Fig. 5, the initial $f_{br}$ has obvious
fluctuation and a slow red-drift before disappearing in the RF spectrum.
Nevertheless, once $f_{br}$ is injection-locked by $f_{mod}$, even
pumped with a free-running laser, no fluctuation and drift are observed
in the RF spectrum. In other words, the breather soliton with locked
frequency is robust with unstable pump condition and the phase noise
of the system is remarkably suppressed much narrower linewidth
of $f_{br}$ beatnote.

\subsection{Controlling of breathing frequency}

Furthermore, we study the locking range of this injection locking
scheme. $f_{mod}$, which is scanned around the initial $f_{br}$, is applied
on the pump laser with modulation power of $0.1\:\mathrm{mW}$ . Figure
6a shows the evolution of the RF spectrum centered at $262\:\mathrm{MHz}$
as we slowly vary $f_{mod}$. In the beginning, when the frequency
difference $\triangle f$ between $f_{mod}$ and initial $f_{br}$
is relatively large, beatnotes of initial $f_{br}$ and $f_{mod}$
and their harmonics components all exist in the RF spectrum. By continuing
the scanning, when $\triangle f$ is less than $\sim15\:\mathrm{MHz}$,
beatnote of initial $f_{br}$ and harmonics components are eliminated,
corresponding to the injection locking of $f_{br}$. Scanning in the same
direction further leads to $f_{mod}$ across the initial $f_{br}$
and out of the locking range at last. We find that after scanning
out of the range, the unlocked $f_{br}$ drifts to a lower frequency
of about $\sim5\:\mathrm{MHz}$, which can be attributed to frequency
red-drift of free-running pump laser during our measurement.

For the purpose of exploring the locking range as a function of modulation
power, the above measurement of locking range is repeated with varied
modulation power and obtained results are depicted in Fig. 6b. With
the increase of the modulation power, the locking range rises rapidly
at the very beginning and then this trend tends to slow down when
the modulation power is beyond $0.2\:\mathrm{mW}$. In our experiment,
the maximum value of modulation power allowing the existence of a
breather soliton state is $1\:\mathrm{mW}$ and the corresponding
locking range is $50\:\mathrm{MHz}$. In the case of modulation power
greater than $1\:\mathrm{mW}$, the breather soliton is annihilated.
We attribute this observation to the excessively strong oscillation
amplitude of the intracavity field induced by a strong modulation
signal, which breaks the condition of the existence of breather soliton
inside the microresonator. This phenomenon indicates that the modulation
power exists an upper limit for effective injection locking.

\begin{figure}
\centerline{\includegraphics[clip,width=1\columnwidth]{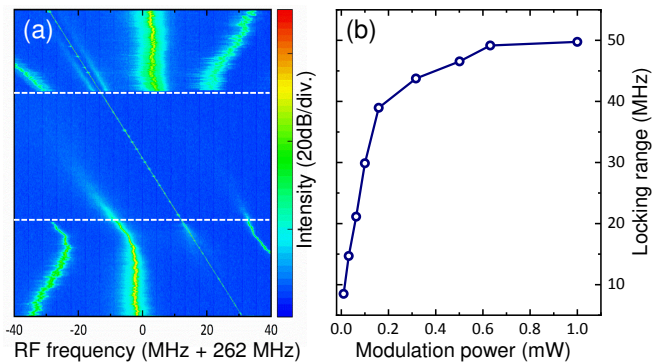}}\caption{\label{fig:Fig6}(a) The evolution of the RF spectrum centered at
$262\:\mathrm{MHz}$ with varied $f_{mod}$. The modulation power
is $0.1\:\mathrm{mW}$ and $f_{br}$ is synchronized to $f_{mod}$
when the frequency difference $\triangle f$ is less than $\sim15\:\mathrm{MHz}$.
(b) Locking ranges with varied modulation power.}
\vspace{-6pt}
\end{figure}

\section{conclusion}

In conclusion, we have experimentally demonstrated the stabilization
and tuning of breathing soliton via injection locking in high-Q $\mathrm{Si_{3}N_{4}}$
microring. By applying an appropriate external modulation signal to
the pump power, the breathing frequency can be effectively locked
to the modulation frequency. Both frequency fluctuation and long term
drift disappear even with a free-running pump laser, showing the high
stability of the sub-GHz breathing frequency. Besides, the linewidth
of breathing frequency beatnote is also become obviously narrower,
indicating the apparent suppression of system phase noise. Being proportional
to the modulation power, the locking range is able to reach $\sim50\:\mathrm{MHz}$
in our experiment. Therefore, a tunable and stable breathing frequency
can be realized by using an EOM or other kinds of modulation instruments
with no need for other feedback stabilization technique. The mechanism
of injection-locking not only is universal for DKS in other material
and structures, but also can be extended to stabilize oscillations
in other nonlinear systems. In addition to the interesting nonlinear
physics underlying the soliton and injection locking, our demonstration
provides a robust way to transfer RF frequencies to broadband optical
frequencies and a potential application in spectroscopy by utilizing
the breathing sidebands to extract the high order spectral information
around each comb line and boost the spectral resolution.

\section*{Experimental section}

Methods and any associated references are mentioned in the Supporting
Information.

\section*{Supporting Information}

Supporting Information is available from the Wiley Online Library
or from the author.

\begin{acknowledgement}
S. Wan and R. Niu contribute equally to this work. The work was supported
by the National Key R\&D Program of China (Grant No.2016YFA0301303),
the National Natural Science Foundation of China (Grant No.11934012
and 11722436), Anhui Initiative in Quantum Information Technologies
(AHY130200), the Fundamental Research Funds for the Central Universities.
This work was partially carried out at the USTC Center for Micro and
Nanoscale Research and Fabrication.
\end{acknowledgement}


%

\bibliographystyle{plainnat}

\end{document}